\documentclass[aps]{revtex4}

\begin{document}

\title{Invariants of the electromagnetic field}
\author{C. A. Escobar$^{1}$ and L. F. Urrutia}
\affiliation{ Instituto de Ciencias Nucleares, Universidad Nacional Aut{\'o}noma de
M{\'e}xico, \\
A. Postal 70-543, 04510 M{\'e}xico D.F., M{\'e}xico}
\affiliation{ Facultad de F\'{\i}sica, Pontificia Universidad Cat\'{o}lica de
Chile, Casilla 306, Santiago 22, Chile}

\begin{abstract}
We present a constructive proof that, in electrodynamics,  all of the gauge-invariant Lorentz scalars and
pseudoscalars can be expressed as functions of the
quadratic ones.
\end{abstract}

\maketitle

\section{Introduction}

In the limit where the gravitational, weak and strong interactions are not
relevant, all classical electric and magnetic phenomena among electrically charged particles can be understood in
terms of Maxwell's equations
\begin{eqnarray}
&&\nabla \cdot \mathbf{B}=0,\qquad \nabla \times \mathbf{E}+\frac{1}{c}\frac{%
\partial \mathbf{B}}{\partial t}=0,\;\;  \label{MAXEQ} \\
&& \nabla \cdot \mathbf{E}=4 \pi \rho, \qquad \nabla \times \mathbf{B}-%
\frac{1}{c}\frac{\partial \mathbf{E}}{\partial t}=\frac{4 \pi}{c}\mathbf{J},
\label{MAXEQ1}
\end{eqnarray}
plus Newton's equations together with the Lorentz force
\begin{equation}
{\mathbf F}= q \left({\mathbf E} + \frac{\mathbf v}{c}\times {\mathbf B}\right). 
\end{equation}
 Here $\mathbf{E}=(E_{1},E_{2},E_{3})$ is the electric field, $%
\mathbf{B}=(B_{1},B_{2},B_{3})$ is the magnetic induction, $\rho$ is
the charge density, $\mathbf{J}$ is the current density,  ${\mathbf v}$ is the velocity of the particle having charge $q$ and $c$ is the invariant speed of light. The above set
of equations yield charge conservation in the form of $\partial \rho
/\partial t+\nabla \cdot \mathbf{J=0.}$ We are using the notation,
conventions and Gaussian units according to Ref. 1. 

The reformulation of Maxwell's equations, Eqs. (\ref{MAXEQ}) and (\ref{MAXEQ1}), in terms
of Lorentz tensors in a four-dimensional Minkowski space (with metric
$\eta _{\mu \nu }=diag(+,-,-,-)$) is much more than a simple matter of
changing to a convenient notation. This is because it makes conceptually transparent the deep
interplay that exists among electric and magnetic phenomena. In the framework of special relativity an event is labeled by the spatial
coordinates $x^{i}\;(i=1,2,3)$ and the time $x^{0}=ct$ (scaled, for dimensional reasons, by the
invariant velocity of light), which are unified
into the contravariant vector  $x^{\mu },\;\mu =0,1,2,3$, that transformes as
\begin{equation}
x^{\prime \mu }=\Lambda _{\;\;\nu }^{\mu }x^{\nu },\qquad \Lambda
_{\;\;\alpha }^{\mu }\eta _{\mu \nu }\Lambda _{\;\;\beta }^{\nu }=\eta
_{\alpha \beta },  \label{LT+VT}
\end{equation}%
when going from an inertial frame $S$ to an inertial frame $S^\prime$. The
coordinate-independent $4\times 4$ matrix $\Lambda _{\;\;\nu }^{\mu }$
describes the Lorentz transformation (rotations and boosts) that relates the two
frames.  The unification of
electricity and magnetism into ED proceeds in an analogous way by recognizing
that, in every inertial frame, the electric field $\mathbf{E}$ and the magnetic induction \ $\mathbf{B}
$ are the components of an antisymmetric two-index tensor $F^{\mu \nu }$,
called the field strength, according to the following
\begin{equation}
F^{0i}=-E_{i},\;\;\;F^{ij}=-\epsilon ^{ijk}B_{k}.
\end{equation}%
Here $\epsilon ^{ijk}\;$is the completely
antisymmetric three-index tensor with $\epsilon ^{123}=+1$. The sources
(charges $\rho $ and currents $\mathbf{J}$) also get unified in the current
four-vector $
J^{\mu }=(c\rho ,\;\mathbf{J}).
$
The transformation properties of the tensor $F^{\mu\nu}$ allows us to determine  how the components $%
E_{i}^{\prime },B_{j}^{\prime }$ look in terms of $E_{i},B_{j}$ when going
from $S$ to $S^{\prime}$.
With the above conventions we can directly  verify that Maxwell's
equations, Eqs. (\ref{MAXEQ}) and (\ref{MAXEQ1}), can be compactly written as
\begin{equation}
\epsilon ^{\mu \nu \rho \sigma }\partial _{\nu }F_{\rho \sigma
}=0,\qquad \partial _{\mu }F^{\mu \nu }=\frac{4\pi }{c}J^{\nu },\;\;
\label{CMAXEQ1}
\end{equation}%
respectively.
The tensor \ $\epsilon ^{\alpha \beta \mu \nu }$  is  completely
antisymmetric in each pair of indices with the convention that $\epsilon ^{0123}=+1$.
The introduction of  the standard potentials $\Phi ,\;\mathbf{A}$, which are unified now in 
four-potential $A_{\mu }=(\Phi ,-\mathbf{A})$ satisfying $F_{\mu \nu
}=\partial _{\mu }A_{\nu }-\partial _{\nu }A_{\mu }$, solves
 the first relation  in (\ref{CMAXEQ1}). Notice also that the gauge
transformations are written now as $A_{\mu }\rightarrow A_{\mu }+\partial
_{\mu }\Theta (x^{\alpha })$, which leave the field strength invariant.

As usual, a covariant index $\mu$ is  obtained by lowering a contravariant index $\nu$ with the metric $\eta_{\mu\nu}$.
Contractions over covariant and a contravariant indexes preserve the tensorial character of the remaining
expression. A notable class of tensors occurs when, after the contractions, there are no free indices left.
In this case we have what is called a scalar quantity under Lorentz
transformations, with the property of having the same value in every
inertial frame. The concept of tensors under Lorentz transformations is generalized by
defining pseudo-tensors as quantities whose transformation rule include an
additional factor of $\det \left[ \Lambda _{\;\;\nu }^{\mu }\right]$. 
Contractions upon a pseudotensor produce another pseudotensor, and
ultimately a pseudoscalar when there are no free indices left.

The tensorial character of Eqs. (\ref{CMAXEQ1}) guarantees that
they adopt the same form in every inertial frame. Also, the tensor
formulation of ED together with the knowledge of the corresponding field
invariants is an essential ingredient in the construction of non linear
extensions of the theory, as well as in the construction of the couplings of
the electromagnetic field to the other basic interactions. These new
interaction terms are encoded in extended gauge invariant Lagrangian densities, which are
required to be Lorentz scalars constructed from the basic fields under
consideration, among which $A_{\mu }$ and/or $F_{\mu \nu }$ will be included.

\section{The field invariants}

\label{S1} This  note provides an explicit
proof of the well-known statement that \textit{in electrodynamics all the
gauge invariant Lorentz scalars and pseudoscalars that can be constructed from the tensors $%
F_{\mu \nu },\eta _{\mu \nu }$\ (or equivalently $\delta _{\nu }^{\mu }$)
and $\epsilon _{\alpha \beta \mu \nu }$ can be ultimately expressed in terms
of the quadratic ones}
\begin{equation}
F=F_{\mu \nu }F^{\mu \nu },\qquad G=F^{\alpha \beta }\tilde{F}_{\alpha \beta
},  \label{QI}
\end{equation}%
where $\tilde{F}_{\alpha \beta }$ is the dual tensor satisfying%
\begin{equation}
\tilde{F}_{\alpha \beta }=\frac{1}{2}\epsilon _{\alpha \beta \mu
\nu }F^{\mu \nu }, \qquad \frac{1}{2}\epsilon _{\alpha \beta \mu \nu }\tilde{F}^{\mu \nu }=-F_{\alpha
\beta }. \label{DUAL}
\end{equation}%
The fact that the only quadratic invariants are those in Eq. (\ref{QI}) is
proved in many text-books.${}^{1-5}$ 

Surprisingly, a search that included well-known graduate text-books in electrodynamics,
produced no examples of a complete proof of the
above statement. Among the many references related to
non-linear electrodynamics, which correctly assume that the most general
Lorentz-invariant Lagrangian that can be written is of the form $L(F,G)$, we
found Ref.6, 
which gives some hints regarding a general proof.
The difficulty in finding a detailed proof of this statement has provided
the motivation for writing this note.

Before going into the details it will be useful to distinguish between two
approaches to the full proof. The first approach, which we call an
existence proof, proceeds along the following lines: since all invariants are
combinations of $F_{\mu\nu}$, $\eta_{\mu\nu}$ and ${\epsilon_{\alpha\beta\mu%
\nu}}$ and we know how to reduce an even product of epsilons, the only
remaining objects that we must encounter at the end of the complete
reduction must be the two quadratic invariants. A similar argument can be
given in terms of the $SO(3)$ vectors $E_i, B_i$, which are the components of ${%
F_{\mu\nu}}$, provided that one first realizes that the only Lorentz invariants
that can be constructed from the  SO(3) quadratical invariants ${\mathbf{E}}^2$, ${\mathbf B
}^2$, ${\mathbf E} \cdot {\mathbf B}$ , are ${\mathbf E}^2-{\mathbf B}^2$ and ${\mathbf E} \cdot {\mathbf B}$.${}^{7}$ 

The
second approach, which we call a constructive proof, classifies all
possible invariants and provides the corresponding algorithm to reduce each
of them to the final functions of $F$ and $G$.

Let us emphasize that our proof is neither the shortest, nor the most
elegant, but it has the following two virtues: (i) it is more than an
existence proof, since it contains a constructive procedure in the sense described
above and (ii) it can be followed and understood by anyone familiar with
standard graduate-level  electrodynamics such as  at the level of Refs.1,2,3,4,5.
with no additional
knowledge of more advanced mathematical tools. For example, a shorter
existence proof can be carried out by using two-component spinor notation
together with the decomposition of $F_{\mu\nu}$ into its selfdual and
anti-selfdual parts.${}^{8}$ 
Another existence proof can be based upon
the Cayley-Sylvester theorem, that determines the number of independent
eigenvalues of $F_{\mu\nu}$, which are calculated in Ref. 9
and
shown to depend only upon $F$ and $G$.${}^{10}$

We now will begin our construction. It is convenient to introduce the following  matrix
notation
\begin{eqnarray}
&&\mathbf{F} \mathbf{=}\left[ F_{\;\;\nu }^{\mu }\right] ,\quad \mathbf{\tilde{F}%
=}\left[ \tilde{F}_{\;\;\nu }^{\mu }\right] ,\quad \mathbf{F}^{p}=\underbrace{%
\mathbf{FF}...\mathbf{F}}_{p\,times}, \quad  \mathbf{F}^{0}=\mathbf{I,}\; \label{MATRIXNOT}\\
&&\mathbf{f} \mathbf{=}\left[ F_{\;\;}^{\mu \nu }\right], \quad \mathbf{f}%
^{T}=-\mathbf{f,\;\;\;\;}\;\;\mathbf{m=}\left[ \eta _{\alpha \beta }\right],
\quad \mathbf{m}^{T}=\mathbf{m},
\end{eqnarray}%
which imply $\mathbf{F=fm}$. In the above, $\mathbf{I}$ is the $4\times 4$ identity operator. Notice that $Tr(\mathbf{F}%
^{2})=-F$. In terms of the matrices with definite symmetry, the product $%
\mathbf{F}^{p}$ can be written as
\begin{equation}
\mathbf{F}^{p}=\underbrace{\left( \mathbf{fm}\right) \left( \mathbf{fm}%
\right) ...\left( \mathbf{fm}\right) }_{p\,times}
\end{equation}%
in such a way that
\begin{equation}
\left( \mathbf{F}^{p}\right) ^{T}=(-1)^{p}\underbrace{\left( \mathbf{mf\;}%
\right) \left( \mathbf{mf\;}\right) ...\left( \mathbf{mf\;}\right) }%
_{p\,times}.  \label{TRANS}
\end{equation}

A convenient way of classifying all the scalars and pseudoscalars is by writing an invariant
of order $n$ (even or odd) in the field strength as
\begin{equation}
I^{(n)}=\underbrace{F^{\mu \nu }F^{\alpha \beta }...F^{\kappa \lambda }}%
_{n\,times}\,I_{\mu \nu \alpha \beta ...\kappa \lambda },
\end{equation}%
where $I_{\mu \nu \alpha \beta ...\kappa \lambda }$ is constructed from the
only tensor and pseudotensor that are invariant  under the proper Lorentz transformations: $\eta
_{\mu \nu }$ and $\epsilon _{\alpha \beta \mu \nu }$. The proof proceeds in
three steps, according to the number of epsilon factors that occur in $%
I_{\mu \nu \alpha \beta ...\kappa \lambda }$, and it is presented in
Sections \ref{S3}, \ref{S4} and \ref{S5}. The appendix includes the basic recurrence relations that are required in the following discussion.

\section{The case of no epsilon factors in $I_{\protect\mu \protect\nu
\protect\alpha \protect\beta ...\protect\kappa \protect\lambda }$}

\label{S3} Since  $I_{\mu \nu \alpha \beta ...\kappa \lambda }$ is now
constructed  from the metric tensor only, a generic member of this class will
have the form
\begin{equation}
Tr(\mathbf{F}^{p})Tr(\mathbf{F}^{q})....Tr(\mathbf{F}^{r}),
\end{equation}%
with $p+q+...+r =n$, and  where we have omitted contractions of the type $\eta
_{\rho \sigma }F^{\rho \sigma }$ that have a zero contribution. In this way,
for this case it is enough to consider the reduction of an arbitrary factor $%
Tr(\mathbf{F}^{p})$.

The antisymmetry of $\mathbf{f}$ guarantees that $Tr(\mathbf{F}^{n})=0,$ for odd
$n$. In fact, from Eq. (\ref{TRANS}) we have
\begin{equation}
Tr(\mathbf{F}^{n})=Tr(\left( \mathbf{F}^{n}\right) ^{T})=\underbrace{-Tr%
\left[ \left( \mathbf{mf\;}\right) \left( \mathbf{mf\;}\right) ...\left(
\mathbf{mf\;}\right) \right] }_{n\,times}=\underbrace{-Tr\left[ \left(
\mathbf{fm}\right) \mathbf{\;}\left( \mathbf{fm}\right) ...\left( \mathbf{fm}%
\right) \right] }_{n\,times}=-Tr(\mathbf{F}^{n})=0,
\end{equation}%
by using the cyclic property of the trace.

The case of an even  $p$ can be reduced by using a recurrence relation, Eq. (\ref%
{REC1}), which ultimately produces
\begin{eqnarray}
Tr(\mathbf{F}^{p}) &=&-\frac{F}{2}Tr(\mathbf{F}^{p-2})+\frac{G^{2}}{16}Tr(%
\mathbf{F}^{p-4}),  \nonumber \\
&& \dots \dots,  \nonumber \\
&=&U(F,G^{2})Tr(\mathbf{F}^{2})+V(F,G^{2})Tr(\mathbf{F}^{0}),\;\;\;  \nonumber
\\
&=&-U(F,G^{2})F+4V(F,G^{2}),
\end{eqnarray}%
where $U,V$ denote the functions of the quadratic invariants that appear in
the reduction process. When similar functions appear below, an analogous  notation will be used
without stating this fact at each step.

In this way, repeated use of Eq. (\ref{REC1}) allows one  to reduce $Tr(\mathbf{F}%
^{p})$ to a function of  $F$ and $G$, where $G$ appears only in even
powers, which is due to parity conservation.

\section{The case of an even number of epsilon factors in $I_{\protect\mu
\protect\nu \protect\alpha \protect\beta ...\protect\kappa \protect\lambda }$%
}

\label{S4}

Using Eq. (\ref{epsi}) as many times as required, we first reduce
all pairs of epsilons, independently of how they are contracted,
and in this  way  this case is transformed into that discussed in  Section \ref{S3}.

\section{The case of an odd number of epsilon factors in $I_{\protect\mu
\protect\nu \protect\alpha \protect\beta ...\protect\kappa \protect\lambda }$%
}

\label{S5}

The use of Eq. (\ref{epsi})\ allows us to reduce this case to the
situation where $I_{\mu \nu \alpha \beta ...\kappa \lambda }$ has only one
epsilon factor, and all the remaining pairs have been written in terms of products of
the metric tensor. We then need to consider only  a generic invariant of
order $n$ with
\begin{equation}
I^{(n)}_{\mu \nu ...\alpha \beta \kappa \lambda \pi \delta }=\underbrace{\eta
_{\mu \nu }...\eta _{\alpha \beta }}_{n-2\,times}\epsilon _{\kappa \lambda
\pi \delta }.
\end{equation}%
Since the indices in $F^{\alpha \beta }$ come in pairs the only form of an
invariant containing only one epsilon factor is
\begin{equation}
I^{(n)}=Tr(\mathbf{F}^{p})\left( \mathbf{F}^{q}\right) ^{\kappa \lambda
}\left( \mathbf{F}^{r}\right) ^{\pi \delta }\epsilon _{\kappa \lambda \pi
\delta },
\end{equation}%
where $p+q+r= n$. The trace part is reduced according to Section \ref{S3}
and we only need to consider the remaining invariant
\begin{equation}
I^{(q+r)}=\left( \mathbf{F}^{q}\right) ^{\kappa \lambda }\epsilon _{\kappa
\lambda \pi \delta }\left( \mathbf{F}^{r}\right) ^{\pi \delta }.
\label{IQPR}
\end{equation}%
Here the reduction is performed by using the recurrence relations of Eqs. (\ref{REC1}), (%
\ref{REC1FEVEN}), and (\ref{REC1FODD}) for each factor of $\left( \mathbf{F}%
^{m}\right) ^{\kappa \lambda }$ that appears  Eq. (\ref{IQPR}). The
intermediate results are
\begin{eqnarray}
\mathbf{F}^{m} &=&W(F,G^{2})\mathbf{I}+Z(F,G^{2})\mathbf{F}^{2},\quad m\,\,%
\mathrm{even}, \\
\mathbf{F}^{m} &=&\bar{Z}(F,G^{2})\mathbf{F}+\bar{W}(F,G^{2})G\mathbf{\tilde{%
F}},\quad m\,\,\mathrm{odd},
\end{eqnarray}%
which yield the final four possibilities
\begin{eqnarray}
&&I_{q\;even,\;r\;even}^{(q+r)}=\left( \mathbf{F}^{2}\right) ^{\kappa
\lambda }\epsilon _{\kappa \lambda \pi \delta }\left( \mathbf{F}^{2}\right)
^{\pi \delta }\;\bar{Z}_{1},  \label{EE} \\
&&I_{q\;even,\;r\;odd}^{(q+r)}=\left( \mathbf{F}^{2}\right) ^{\kappa \lambda
}\epsilon _{\kappa \lambda \pi \delta }{F}^{\pi \delta }\;\bar{Z}_{2}+\left(
\mathbf{F}^{2}\right) ^{\kappa \lambda }\epsilon _{\kappa \lambda \pi \delta
}\tilde{F}^{\pi \delta }\;\bar{Z}_{3},  \label{EO} \\
&&I_{q\;odd,\;r\;even}^{(q+r)}=F^{\kappa \lambda }\epsilon _{\kappa \lambda
\pi \delta }\left( \mathbf{F}^{2}\right) ^{\pi \delta }\;\bar{Z}_{4}+\tilde{F%
}^{\kappa \lambda }\epsilon _{\kappa \lambda \pi \delta }\left( \mathbf{F}%
^{2}\right) ^{\pi \delta }\;\bar{Z}_{5},  \label{OE} \\
&&I_{q\;odd,\;r\;odd}^{(q+r)}=\tilde{F}^{\kappa \lambda }\epsilon _{\kappa
\lambda \pi \delta }\tilde{F}^{\pi \delta }\;\bar{Z}_{6}+\tilde{F}^{\kappa
\lambda }\epsilon _{\kappa \lambda \pi \delta }{F}^{\pi \delta }\;\bar{Z}%
_{7}+F^{\kappa \lambda }\epsilon _{\kappa \lambda \pi \delta }F^{\pi \delta
}\;\bar{Z}_{8},  \label{OO1}
\end{eqnarray}%
where $\bar{Z}_{A},\;A=1,2,3,...,8$ are already functions of $F$ and $G$. \
The final reduction of the terms involved above proceeds as follows. Equation (\ref%
{EE}) yields
\begin{eqnarray}
\left( \mathbf{F}^{2}\right) ^{\kappa \lambda }\epsilon _{\kappa \lambda \pi
\delta }\left( \mathbf{F}^{2}\right) ^{\pi \delta }=0,
\end{eqnarray}%
because $\left( \mathbf{F}^{2}\right) ^{\kappa \lambda }=\left( \mathbf{F}%
^{2}\right)^{\lambda \kappa}$. The reduction of Eqs. (\ref{EO}) and (\ref{OE}%
) include the same terms and have  two contributions that yield
\begin{eqnarray}
&& \left( \mathbf{F}^{2}\right) ^{\kappa \lambda } \epsilon _{\kappa \lambda
\pi \delta }F^{\pi \delta }=0, \qquad \tilde{F}^{\kappa \lambda }\epsilon
_{\kappa \lambda \pi \delta }\left( \mathbf{F}^{2}\right) ^{\pi \delta }=0,
\end{eqnarray}
again due to the symmetric character of $\left( \mathbf{F}^{2}\right)
^{\kappa \lambda }$. The last reduction arises from Eq.(\ref{OO1}) which
includes
\begin{eqnarray}
\tilde{F}^{\kappa \lambda }\epsilon _{\kappa \lambda \pi \delta }\tilde{F}%
^{\pi \delta }=-2G \qquad \tilde{F}^{\kappa \lambda }\epsilon _{\kappa
\lambda \pi \delta }F^{\pi \delta }=-2F \qquad F^{\kappa \lambda }\epsilon
_{\kappa \lambda \pi \delta }F^{\pi \delta } =2G.
\end{eqnarray}
Here we have made use of the second relation in Eq. (\ref{DUAL}) together with the definition of $G$.
This completes the proof.

\section*{ACKNOWLEDGEMENTS}
L.F.U is partially supported by the project DGPA-IN109013 and a
sabbatical fellowship from DGAPA-UNAM. He also acknowledges the hospitality
of the Facultad de F\'\i sica and  support from the Programa de
Profesores Visitantes, at the Pontificia Universidad Cat\'olica de Chile.
C.A.E acknowledges support from a CONACyT graduate fellowship as well as
partial support from the project DGAPA-IN109013 and the program PAEP at
UNAM. The authors gratefully acknowledge Dr. E. Nahmad  and Dr. M. A. Garc\'\i a
for useful suggestions.

\appendix
\section{}
\label{APP}
The basic relations employed in the proof are
\begin{equation}
F_{\;\;\lambda }^{\mu }F_{\;\;\nu }^{\lambda }=\tilde{F}_{\;\;\lambda }^{\mu
}\tilde{F}_{\;\;\nu }^{\lambda }-\frac{1}{2}F\delta _{\nu }^{\mu },
\label{REL1}
\end{equation}%
\begin{equation}
\tilde{F}_{\;\;\lambda }^{\mu }F_{\;\;\nu }^{\lambda }=-\frac{1}{4}G\;\delta
_{\nu }^{\mu },  \label{REL2}
\end{equation}%
which are well known.${}^{6,9,11,12}$ 
The first one is a
direct consequence of the definition of the dual tensor together with the
property%
\begin{equation}
\epsilon _{\mu \nu \rho \sigma }\epsilon _{\pi \delta \kappa \lambda }=\det
\left[
\begin{array}{cccc}
{\eta}_{\mu \pi } & {\eta}_{\mu \delta } & {\eta}_{\mu \kappa } & {\eta}%
_{\mu \lambda } \\
{\eta}_{\nu \pi } & {\eta}_{\nu \delta } & {\eta}_{\nu \kappa } & {\eta}%
_{\nu \lambda } \\
{\eta}_{\rho \pi } & {\eta}_{\rho \delta } & {\eta}_{\rho \kappa } & {\eta}%
_{\rho \lambda } \\
{\eta}_{\sigma \pi } & {\eta}_{\sigma \delta } & {\eta}_{\sigma \kappa } & {%
\eta}_{\sigma \lambda }%
\end{array}%
\right] .  \label{epsi}
\end{equation}%
Equation (\ref{REL2}) arises by parity arguments and the fact that the only
quadratic invariant that violates parity is $G$. In this way, the Lorentz covariance
demands that
\begin{equation}
\tilde{F}_{\;\;\lambda }^{\mu }F_{\;\;\nu }^{\lambda }=\alpha \;G\;\delta
_{\nu }^{\mu },  \label{AP1}
\end{equation}%
where the factor $\alpha =-1/4$ is obtained by contracting  Eq. (\ref%
{AP1}) with $\delta _{\mu }^{\nu }$ and using the definition of $G$.

It is now
convenient to go back to the matrix notation introduced in Eq.(\ref{MATRIXNOT}) and rewrite Eqs. (\ref{REL1}) and (\ref{REL2}) as 
\begin{equation}
\mathbf{F}^{2}=\mathbf{\tilde{F}}^{2}-\frac{1}{2}F\;\mathbf{I,}
\label{REL11}
\end{equation}%
\begin{equation}
\mathbf{\tilde{F}F=F\tilde{F}}=-\frac{1}{4}G\;\mathbf{I}.  \label{REL21}
\end{equation}%
 The above equations allow us to
write the following recursion relation
\begin{equation}
\mathbf{F}^{p}=-\frac{F}{2}\mathbf{F}^{p-2}+\frac{G^{2}}{16}\mathbf{F}%
^{p-4},\;  \label{REC1}
\end{equation}%
which is obtained as follows%
\begin{eqnarray}
\mathbf{F}^{p} &=&\mathbf{F}^{p-2}\mathbf{F}^{2}=\mathbf{F}^{p-2}\mathbf{%
\tilde{F}}^{2}-\frac{1}{2}F\;\mathbf{F}^{p-2},  \nonumber \\
\mathbf{F}^{p} &=&\mathbf{F}^{p-4}\mathbf{FF\tilde{F}\tilde{F}}-\frac{1}{2}%
F\;\mathbf{F}^{p-2},  \nonumber \\
\mathbf{F}^{p} &=&\frac{1}{16}G^{2}\mathbf{F}^{p-4}-\frac{1}{2}F\;\mathbf{F}%
^{p-2},  \label{RED1}
\end{eqnarray}%
where we have used Eq. (\ref{REL11}) in the first line, together with Eq.(%
\ref{REL21}) twice in the second line. For an even $p$, the recurrence will
end at $p=4$, leading to
\begin{equation}
\mathbf{F}^{4}=\frac{1}{16}G^{2}\mathbf{I}-\frac{1}{2}F\;\mathbf{F}^{2}.
\label{REC1FEVEN}
\end{equation}%
In the case of an odd $p$, the final result will correspond to $p=3$
\begin{equation}
\mathbf{F}^{3}=\mathbf{F}\left( \mathbf{F}^{2}\right) =\mathbf{F}\left(
\mathbf{\tilde{F}\tilde{F}}-\frac{1}{2}F\mathbf{I}\right) =-\frac{1}{4}G%
\mathbf{\tilde{F}}-\frac{1}{2}F\mathbf{F}.  \label{REC1FODD}
\end{equation}
In an analogous way we can show that
\begin{equation}
\mathbf{\tilde{F}}^{p}=\frac{1}{16}G^{2}\mathbf{\tilde{F}}^{p-4}+\frac{1}{2}%
F\;\mathbf{\tilde{F}}^{p-2}.  \label{REC2}
\end{equation}%
Again, some care must be taken in the final step of the above recurrence relation. We
obtain%
\begin{eqnarray}
&& p \,\, \mathrm{even}: \qquad  \mathbf{\tilde{F}}%
^{4}=\left( \frac{1}{16}G^{2}+\frac{1}{4}F^{2}\right) \mathbf{I} + \frac{1}{2%
}F\mathbf{F}^{2} ,  \label{REC2FEVEN} \\
&& p \,\,\,\, \mathrm{odd}: \qquad  \mathbf{\tilde{F}}^{3}=-\frac{%
1}{4}G \mathbf{F} + \frac{1}{2}F\mathbf{\tilde{F}}.  \label{REC2FODD}
\end{eqnarray}
Other useful relations that can be derived directly from the above are
\begin{eqnarray}
&&\hspace{3cm} \tilde{F}^{\alpha \beta }\epsilon _{\beta \mu \nu \rho }=\left( F_{\mu \nu
}\delta _{\rho }^{\alpha }+F_{\nu \rho }\delta _{\mu }^{\alpha }+F_{\rho \mu
}\delta _{\nu }^{\alpha }\right), \\
&&\mathbf{F}^{m}\mathbf{\tilde{F}}^{n}=\left( -\frac{G}{4}\right) ^{n}\mathbf{F%
}^{m-n},\,\, m > n, \quad \quad 
\mathbf{F}^{m}\mathbf{\tilde{F}}^{n}=\left( -\frac{G}{4}\right) ^{m}\mathbf{%
\tilde{F}}^{n-m},\,\, m < n.
\end{eqnarray}

\

\

\end{document}